\documentclass[aps,pra,superscriptaddress,twocolumn,10pt]{revtex4-1}

\usepackage{amsmath,amssymb,mathtools}
\usepackage[per-mode=symbol, separate-uncertainty = true, multi-part-units = single]{siunitx}
\usepackage{xspace}		
\usepackage[hidelinks]{hyperref} 
\usepackage{booktabs}
\usepackage[version=3]{mhchem}
\graphicspath{{./figures/}}
\usepackage[normalem]{ulem}

\let\orgautoref\autoref
\providecommand{\Autoref}{%
  \def\sectionautorefname{Section}%
  \def\figureautorefname{Figure}%
  \def\subfigureautorefname{Figure}%
  \orgautoref}
\renewcommand{\autoref}{%
  \def\sectionautorefname{Sec.}%
  \def\figureautorefname{Fig.}%
  \def\subfigureautorefname{Fig.}%
  \orgautoref}

\newcommand\costhetwod[0]{\langle\cos^2\theta_{2D}\rangle\xspace}

\newcommand\DIB[0]{DIB\xspace}
\newcommand\DIBlong[0]{\mbox{1,4-diiodobenzene}\xspace}

\newcommand\DBBlong[0]{\mbox{1,4-dibromobenzene}\xspace}

\usepackage{color}

\newcommand\panelref[2]{\hyperref[#1]{\autoref*{#1} (#2)}\xspace} 
\newcommand\Panelref[2]{\hyperref[#1]{\Autoref*{#1} (#2)}\xspace} 
\newcommand\panelrefonly[2]{\hyperref[#1]{(#2)}\xspace} 
\newcommand{\degree}{\ensuremath{^\circ}}%

\newcommand{\ket}[1]{\, | #1 \rangle}

\begin{document}


\title{Strongly aligned molecules inside helium droplets in the near-adiabatic regime} 



\author{Benjamin Shepperson}
\affiliation{Department of Chemistry, Aarhus University, Langelandsgade 140, DK-8000 Aarhus C, Denmark}

\author{Adam S. Chatterley}
\affiliation{Department of Chemistry, Aarhus University, Langelandsgade 140, DK-8000 Aarhus C, Denmark}

\author{Anders A. S{\o}ndergaard}
\affiliation{Department of Chemistry, Aarhus University, Langelandsgade 140, DK-8000 Aarhus C, Denmark}

\author{Lars Christiansen}
\affiliation{Department of Chemistry, Aarhus University, Langelandsgade 140, DK-8000 Aarhus C, Denmark}

\author{Mikhail Lemeshko}
\affiliation{IST Austria (Institute of Science and Technology Austria), Am Campus 1, 3400 Klosterneuburg, Austria}

\author{Henrik Stapelfeldt}
\affiliation{Department of Chemistry, Aarhus University, Langelandsgade 140, DK-8000 Aarhus C, Denmark}


\date{\today}

\begin{abstract}
Iodine (\ce{I_2}) molecules embedded in He nanodroplets are aligned by a 160 ps long laser pulse.
The highest degree of alignment, occurring at the peak of the pulse and quantified by $\costhetwod$, is measured as a function of the laser intensity. The results are well described by $\costhetwod$ calculated for a gas of isolated molecules each with an effective rotational constant of 0.6 times the gas-phase value, and at a temperature of 0.4 K. Theoretical analysis using the angulon quasiparticle to describe rotating molecules in superfluid helium rationalizes why the alignment mechanism is similar to that of isolated molecules with an effective rotational constant. A major advantage of molecules in He droplets is that their 0.4 K temperature leads to stronger alignment than what can generally be achieved for gas phase molecules -- here demonstrated by a direct comparison of the droplet results to measurements on a $\sim$ 1 K supersonic beam of isolated molecules. This point is further illustrated for more complex system by measurements on \DIBlong and \DBBlong. For all three molecular species studied the highest values of $\costhetwod$ achieved in He droplets exceed 0.96.

\end{abstract}


\maketitle 

\section{Introduction}

Recent experimental work demonstrates that it is possible to align molecules embedded in helium nanodroplets using laser-induced methods normally applied to gas-phase molecules~\cite{pentlehner_impulsive_2013,pentlehner_laser-induced_2013,christiansen_alignment_2015,Shepperson-2017-nonadiabatic-short}. Here alignment refers to confinement of molecular axes to axes fixed in the laboratory frame -- typically defined by the polarization vectors of the laser pulses employed~\cite{stapelfeldt_colloquium:_2003,LemKreDoyKais13}. The transfer of laser-induced alignment techniques from gas-phase molecules to molecules in He droplets opens new opportunities such as extending the range of molecular systems amenable to alignment -- notably to larger molecules, clusters and complexes~\cite{choi_infrared_2006,yang_helium_2012,florez_ir_2015}, conducting experiments on aligned molecules in the presence of a dissipative environment~\cite{pentlehner_impulsive_2013,hartmann_dissipation_2013,chaussard_dissipation_2015} and possibly using controlled rotation of molecules~\cite{korobenko_direct_2014} to study the properties of He droplets~\cite{gomez_shapes_2014}.

Most studies conducted so far focused on alignment in the impulsive regime where the duration of the alignment laser pulse is much shorter than the rotational period of the molecules under study and showed that the alignment dynamics differs strongly from that of isolated molecules. Most recently a combined experimental and theoretical effort demonstrated that for short, sufficiently weak pulses a molecule and its solvation shell of He atoms~\cite{kwon_quantum_2000} can be set into coherent rotation that lasts long enough to create revival structures in the time-dependent alignment signals~\cite{Shepperson-2017-nonadiabatic-short}. As the intensity of the alignment pulses is increased the frictionless rotation is perturbed leading to a faster loss of coherence causing the revivals to disappear. For even stronger pulses the initial alignment dynamics changes dramatically and resembles that of isolated molecules indicating that the fast induced rotation leads to a transient decoupling of the molecule from the surrounding He atoms.

In the adiabatic limit a single experimental study has been reported in which the degree of alignment was measured as a function of intensity, up to $\sim$ 1 TW/cm$^2$, for three different molecular species~\cite{pentlehner_laser-induced_2013}. In the case of \DIBlong molecules the degree of alignment was similar for molecules in He droplets and isolated molecules in a supersonic molecular beam. For iodobenzene and methyliodide molecules in He droplets the degree of alignment appeared to be significantly lowered than that of the isolated molecules, in particular at the higher intensities of the alignment pulse.

Here we present a more detailed and quantitative investigation of adiabatic alignment of molecules in He droplets which significantly improves our understanding of the process. One key experimental parameter is strongly improved compared to previous work. Namely, we deconvolute the nonaxial recoil effect from the angular distribution of recoiling fragment ions, created by laser-induced Coulomb explosion, used to measure the molecular alignment~\cite{christensen_deconvoluting_2016}. This provides a much more accurate determination of the degree of alignment and allows for a quantitative comparison to theoretical calculations for the case of iodine molecules. The outcome of the comparison strongly indicates that the mechanism of adiabatic alignment for molecules in He droplets is the same as for gas-phase molecules except that the molecules in the droplet possess an effective moment of inertia due to the interaction with the surrounding He atoms~\cite{kwon_quantum_2000}. The improved precision in the alignment determination also reveals a very high degree of alignment both for iodine and the two other molecules studied,  \DIBlong (DIB) and \DBBlong (DBB). Comparison to alignment of isolated molecules under identical laser conditions identifies the strong alignment as being due to the low (0.4 K) temperature of the molecules inside the He droplets.

The paper is organized as follows. In \autoref{sec:exp}, the experimental setup is described. \Autoref{sec:theory} explains the theory based on the angulon quasiparticle. The angulon represents a quantum rotor dressed by a many-particle field of superfluid excitations~\cite{schmidt_rotation_2015, schmidt_deformation_2016}, and was shown to provide a reliable model for a molecule rotating in superfluid helium~\cite{LemeshkoDroplets16, YuliaPhysics17, Shepperson-2017-nonadiabatic-short}. In \autoref{sec:res-iodine}, the experimental results for iodine molecules - both inside droplets and in gas pase are presented and compared to calculations based on a numerical solution of the rotational Schr\"{o}dinger equation for a linear molecule interacting with the alignment laser pulse. \Autoref{sec:res-dib-dbb} presents experimental results for alignment of DIB and DBB and conclusions are drawn in \autoref{sec:conclusion}.

\section{Experimental Setup}\label{sec:exp}
\subsection{Helium droplet apparatus}\label{sec:A}

\begin{figure}%
\includegraphics[width=\columnwidth]{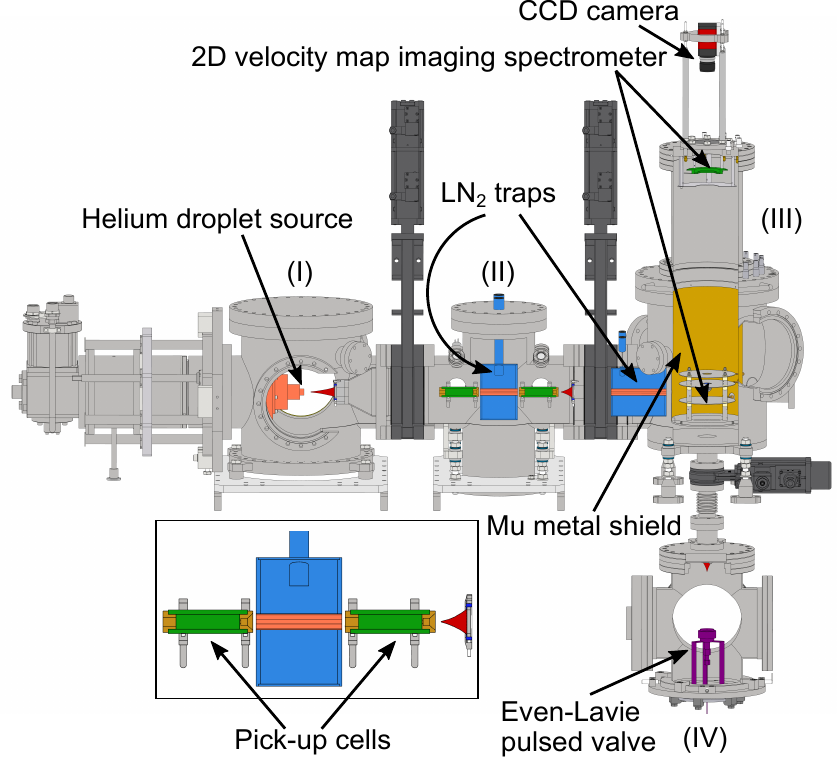}
\caption{Schematic diagram showing the helium droplet machine. Depicted from left to right are the helium droplet source chamber (I), the pick-up chamber (II), and the target chamber (III). Beneath the target chamber is a second source chamber (IV) that houses a pulsed Even-Lavie valve. The target chamber is equipped with a 2D velocity map imaging spectrometer. The design of the pick-up cells and the liquid nitrogen trap are shown in the insert.}
\label{fig:Heliumdropletapparatus}%
\end{figure}

A schematic diagram of the helium droplet machine used in this experiment is shown in \autoref{fig:Heliumdropletapparatus}. The machine consists of four differentially pumped vacuum chambers: the helium droplet source chamber (I), the pick-up chamber (II), the target chamber (III), and another source chamber (IV) housing a pulsed supersonic Even-Lavie valve~\cite{even_cooling_2000}. All of the chambers can be isolated from one another by closing pneumatic gate valves. A continuous liquid helium droplet beam is produced in chamber I by expanding pre-cooled (T \textless 25 K) high purity helium gas (99.9999 $\%$) with sufficiently high stagnation pressure (P \textgreater 10 bar) through a droplet source consisting of a 2-mm-diameter disk with a 5-$\mu$m-diameter aperture (Platinum/Iridium (95/5) electron microscope aperture produced by PLANO). This aperture is compression sealed to a hollow copper assembly using a gold washer. The complete assembly is mounted onto the tip of a closed cycle cryostat (SHI, SRDK-415D). In this study two different droplet sizes were employed, for the iodine experiment the expansion conditions used were 14 K and 25 bar and for DIB and DBB 11 K and 25 bar. Slightly larger droplet sizes were required to pick up \DIBlong molecules due to its lower vapour pressure.

The temperature of the nozzle assembly is controlled by using two high power resistors (Farnell, MHP35 470F) and a Lakeshore 335 temperature controller. The temperature of the droplet source is measured by a silicon diode (LakeShore, DT-670B-CU) that is located close to the 5-$\mu$m aperture. 
To ensure efficient pre-cooling of the helium gas two meters of copper pipe (1/16 inch) is wrapped around both the first and second stage of the cold head. 
To reduce the radiative heat load acting on the cold head the entire nozzle assembly is enclosed within a copper radiation shield that is attached to the first stage of the cold head.

The closed cycle cryostat is mounted on an X,Y,Z translation stage which allows fine alignment of the droplet beam to a 1-mm-diameter skimmer (Beam dynamics, model 50.8) that is positioned at a distance of 1.8 mm downstream from the 5-$\mu$m aperture. After passing through the skimmer the droplet beam enters the pick-up chamber where the vapor of a species of interest can be finely added into a pick-up cell using an ultra-fine needle valve (Kurt Lesker, VZLVM267). In this work the vapor pressure was controlled such that at most a single molecule of either iodine, \DIBlong or \DBBlong was picked up by the droplets. The pick-up chamber contains two pick-up cells in tandem (see insert on \autoref{fig:Heliumdropletapparatus}) which allows for the possibility of doping a helium droplet with two different species. In the studies here only a single cell was used. A liquid nitrogen cooled trap is positioned in-between the two pick-up cells to reduce the amount of effusive species that are not captured by the helium droplets from traveling towards the target chamber. After exiting the pick-up region the droplet beam passes through a 2-mm-skimmer (Beam dynamics, model 2) and enters the target chamber. The droplet beam then passes through a second liquid nitrogen trap (indicated by blue on \autoref{fig:Heliumdropletapparatus}) and finally arrives in the interaction region of a 2D imaging spectrometer.

The source chamber placed beneath the target chamber houses a pulsed Even-Lavie valve that is used for the study of isolated molecules. In these experiments, the molecular beam is formed by expanding high-pressure helium gas (80 bar) mixed with the molecular species through the pulsed valve. Before entering the interaction region of the target chamber the molecular beam passes through a 3-mm-diameter skimmer (Beam dynamics, model 2) positioned at a distance of 13 cm from the exit of the pulsed valve. All the molecules used in this study were placed onto filter paper and inserted into the sample holder located inside the valve. In order to get sufficient ion signal of \DIBlong the valve was heated to 85\degree C. For \DBBlong and iodine no heating was applied. For all three molecular species we estimate the partial vapour pressure to be about 1 mbar. All experiments on isolated molecules were performed at a repetition rate of 100 Hz.

To ensure ultra-high vacuum conditions the entire system is pumped by maglev turbo molecular pumps backed by dry backing pumps. This eliminates the risk of oil contamination, which could otherwise lead to the droplets capturing these impurities. The helium droplet source chamber is pumped by a 2200 L/s vacuum pump (Edwards, STPiXR2206), the pick-up chamber is pumped by a 450 L/s vacuum pump (Edwards, STPiX455) and both the target and molecular beam source chamber are each pumped by a by a 1100 L/s vacuum pump (Edwards, STPiXR1606). Both of the source chambers share the same backing line which is pumped by a 620 m$\textsuperscript{3}$/h dry pump (Edwards, IGX 600 L). The target and pick-up chambers also share a backing line which is pumped by a 100 m$\textsuperscript{3}$/h dry pump (Busch Vacuum, Cobra BA 100).




\subsection{2D imaging spectrometer and laser setup}
\label{sec:B}

The target chamber (III) contains a standard 2D velocity map imaging spectrometer \cite{chandler_twodimensional_1987,eppink_velocity_1997} that consists of an open three electrode electrostatic lens and a 50-mm-diameter microchannel plate (MCP) detector combined with a similar sized phosphor screen. The electrodes are mounted onto the bottom of a mu-metal cylinder that is attached to a CF 250 flange with high voltage feedthroughs for the three electrodes of the electrostatic lens: the repeller, extractor and ground. The MCP detector is mounted onto a separate CF 160 flange with high voltage feedthroughs for the phosphor screen and the front and backside of the MCP. With this configuration it is possible to insert a flight tube that allows for better velocity resolution of low kinetic energy ions.

The interaction region is located in-between the repeller and extractor electrodes. In this region either the droplet beam or the beam of isolated molecules is crossed perpendicularly by two collinear 800 nm pulses both provided by the same Ti:Sapphire laser (Spitfire-ACE-35F, Spectra-Physics). The molecules initially interact with a pulse from the strongly chirped (GDD = 1.7 ps$^2$) uncompressed output of the laser, that is used to align the molecules, hereafter referred to as the alignment pulse. Its duration, $\tau_\text{align}$, is 160 ps (FWHM). The second pulse is taken from the compressed output of the laser and is used to measure the spatial orientation of the molecules through Coulomb explosion. This pulse will be referred to as the probe pulse ($\tau_\text{probe}$ = \SI{40}{fs} (FWHM), I$_\text{probe}$ = $\SI{3.7e14}{W/cm^2}$). Both beams pass through a 30 cm focusing lens prior to entering the target chamber, which gives a focal spot size of $\omega_\text{0}$ = 20 $\mu$m for the probe and $\omega_\text{0}$ = 36 $\mu$m for the alignment beam. The probe pulses are sent at time $t$ with respect to the peak of the alignment pulses. The molecular ions that are formed following Coulomb explosion recoil with directions that are given by the angular distribution of the molecular axes at the instant of the probe pulse. These ions are extracted and impact the MCP causing a discharge of electrons. The ion images are recorded by a charge-coupled-device (CCD) camera (Allied Vision Prosilica GE680) that monitors the phosphor screen and captures the fluorescence created by electrons impacting on the phosphor screen. The MCP is gated in time by a high voltage switch, so that only a single ion mass is detected at a time. Only the individual coordinates of each ion hit are saved and analysis of the recorded data can be done offline.

\section{The angulon theory}
\label{sec:theory}


Recently it has been shown that molecules in superfluid helium droplets can be conveniently described as `angulon' quasiparticles~\cite{LemeshkoDroplets16, YuliaPhysics17, LemSchmidtChapter2017,  Shepperson-2017-nonadiabatic-short, schmidt_rotation_2015, schmidt_deformation_2016,  Yakaboylu16, Bikash16, Li16,  Redchenko16}. Here we use the angulon theory as a simple model to show that under the experimental conditions reported here, the alignment of I$_2$, DIB, and DBB in the presence of helium can be approximated as the gas-phase alignment. The only difference is contained in  the renormalized value of the rotational constant.  The angulon represents a quantum rotor dressed by a many-particle field of helium excitations and is given by the solutions to the following Hamiltonian~\cite{schmidt_rotation_2015}:
\begin{multline}
\label{hamiltonian}
 \widehat{H} = B \hat{\mathbf{J}}^2 + \sum\limits_{k\lambda\mu} \omega_k \hat{b}^\dag_{k\lambda\mu} \hat{b}_{k\lambda\mu}
 \\ + \sum\limits_{k\lambda\mu} U_\lambda(k) ~ [Y^*_{\lambda\mu}(\hat{\theta},\hat{\phi})\hat{b}^\dag_{k\lambda\mu}+Y_{\lambda\mu}(\hat{\theta},\hat{\phi}) \hat{b}_{k\lambda\mu}],
\end{multline}
where we used the notation  $\sum_k \equiv \int dk$, and set $\hbar \equiv 1$. The first term of Eq.~\eqref{hamiltonian} gives the rotational kinetic energy of a linear-rotor molecule, such as I$_2$, where $\hat{\mathbf{J}}$ is the angular momentum operator and $B = 1/(2I)$ is the molecular rotational constant, with $I$ the molecular moment of inertia.  For nonlinear molecules, such as DIB and DBB, the angulon Hamiltonian will assume a different form~\cite{Cherepanov17}, however, Eq.~\eqref{hamiltonian} can still be used to estimate the average effect of the superfluid bath on the molecular rotational structure~\cite{LemeshkoDroplets16}.

The second term of Eq.~\eqref{hamiltonian} corresponds to the kinetic energy of the excitations in superfluid helium, such as phonons and rotons, whose spectrum is given by the dispersion relation $\omega_k$~\cite{Donnelly1998}.  The operators $\hat{b}^\dag_{k\lambda\mu}$ ($\hat{b}_{k\lambda\mu}$) are creating (annihilating) a superfluid excitation with linear momentum $k=|\mathbf{k}|$, the angular momentum $\lambda$, and its projection, $\mu$, onto the $z$-axis~\cite{LemSchmidtChapter2017}.

The last term of the Hamiltonian~\eqref{hamiltonian} gives the exchange of momentum and angular momentum between the molecular impurity and the superfluid. Here the coupling constants $U_\lambda(k)$ are expressed through the corresponding Legendre moments of the molecule--Helium potential energy surface in  Fourier space~\cite{LemSchmidtChapter2017, schmidt_rotation_2015}; the spherical harmonics $Y_{\lambda\mu}(\hat{\theta},\hat{\phi})$~\cite{VarshalovichAngMom} explicitly depend on the molecular angle operators in the laboratory frame.

Following  Ref.~\cite{LemeshkoDroplets16}, we can parametrize the molecule-He interactions, $U_\lambda(k)$ of Eq.~\eqref{hamiltonian}, by a single parameter, $\Delta$, which gives the frontal-to-lateral anisotropy of the molecule-He potential energy surface. Then, the interaction of a molecule with superfluid helium can be quantified by the following dimensionless parameter:
\begin{equation}
\label{gamma}
\gamma = B/\Delta
\end{equation}
The species for which $\gamma < 1$ and $\gamma > 1$ belong to the strong-coupling and weak-coupling regimes, respectively. For I$_2$, $\gamma \approx 0.016 \ll 1$~\cite{LemeshkoDroplets16}, which assures  the applicability of the strong-coupling angulon theory~\cite{schmidt_deformation_2016}. Up to our knowledge, an exact shape of the potential energy surface for the  DIB and DBB molecules interacting  with helium is not available in the literature. However, we can assume that the interaction anisotropy is comparable to that for I$_2$, and the strong-coupling approximation can be therefore applied. 

In such a regime, the field-free molecular rotation can be described as that of a gas-phase molecule with a renormalized rotational constant~\cite{LemeshkoDroplets16}:
\begin{equation}
\label{Bstar}
	 B^\ast = B \left( 1-  \beta  \right)^2,
\end{equation}
where
\begin{equation}
\label{beta}
	\beta= \left( \sum_{k,\lambda} a_\lambda \frac{ U_\lambda(k)^2}{\omega^2_{k}} \right)^{1/2}
\end{equation}
with $a_\lambda =  \lambda(\lambda+1)(2\lambda+1) / (8 \pi)$.
Within our phenomenological approach, we treat $\beta$ as a free parameter and set it to $\beta=0.23$, which reproduces the results of quantum Monte Carlo calculations giving $B^\ast = 0.6 B$~\cite{Shepperson-2017-nonadiabatic-short}.
In the strong-coupling approximation, the angulon eigenstates corresponding to angular momentum $\vert L, M \rangle$ are given by~\cite{schmidt_deformation_2016}:
\begin{equation}
\label{transHground}
	\ket{\psi_{LM}} =  e^{  \sum_{k \lambda}  \frac{U_\lambda (k)}{\omega_{k}} \left[Y_{\lambda\mu}(\hat{\theta},\hat{\phi}) \hat{b}_{k\lambda\mu} - Y^*_{\lambda\mu}(\hat{\theta},\hat{\phi})\hat{b}^\dag_{k\lambda\mu}  \right ]} \ket{0} \ket{LM},
\end{equation}

In the presence of an alignment laser field, two additional parameters -- the laser pulse duration and intensity  -- need to be taken into account. In our treatment we assume that the molecule forms the angulon -- a rotor dressed in a coat of superfluid excitations. In order for the angulon to interact with the laser field adiabatically, the characteristic timescale of  helium excitations, $\tau_\text{He}$ has to be much shorter compared to the alignment laser pulse duration, $\tau_\text{align}$.  The time scale of excitations in superfluid helium  is given by $\tau_{He} \sim \mu^{-1}$, where $\mu \approx 5$~cm$^{-1}$ is the chemical potential (i.e.\ the binding energy per one He atom) of superfluid helium~\cite{SyvokonLTPhys06, toennies_superfluid_2004}. Thus we obtain $\tau_{He} \sim 7$~ps, which is indeed much smaller compared to the pulse duration $\tau_\text{align} = 160$~ps. Following the same logic, the pulse should also be adiabatic with respect to molecule-helium interactions, which is satisfied as well since $\tau_\text{align} \Delta \approx 11  \gg  1$. As long as the two adiabaticity conditions are fulfilled,  at every given moment of time during the pulse, the molecule turns into an angulon at a much  faster timescale compared to the changes in the applied laser field.  Moreover, at the intensity of 1~TW/cm$^{2}$, we estimate the molecule-laser coupling for I$_2$ to be $\sim 64$~cm$^{-1}$, which is much larger than the energy scale of the I$_2$--He interactions, which is given by $\Delta = 2.35$~cm$^{-1}$~\cite{LemeshkoDroplets16}. Therefore, here we encounter the strong-field limit of the angulon, which was previously dubbed the `pendulon'~\cite{Redchenko16}.

 Thus, we can reduce the problem of the molecule interacting with helium and the laser field to the problem of an angulon in a laser field, as given by the Hamiltonian:
\begin{equation}
 \label{eq:AngLaser}
\hat H = \hat H_A - \frac{1}{4}\Delta\alpha E_\text{align}^2 \cos^2\theta,
\end{equation}
where $\Delta\alpha$ is the polarizability anisotropy, $E_\text{align}$ the E-field of the alignment pulse and $\theta$ the polar angle between the molecular axis and the alignment pulse polarization.
Here, in the strong-coupling approximation, the eigenenergies of $H_A$ are given by
\begin{equation}
 \label{eq:AngLaser}
E^{(A)}_{LM} = B^\ast J(J+1),
\end{equation}
 In the calculations of the alignment dynamics, the matrix elements of the $\hat{\mathbf{J}}^2$ and $\cos^2\theta$ operators do not act in the superfluid Hilbert space. Therefore, the corresponding matrix elements between the dressed states~\eqref{transHground} coincide with the gas-phase matrix elements, and all the difference compared to the gas phase is contained in the renormalized value of the rotational constant $B^\ast$.


\section{Results for iodine molecules}
\label{sec:res-iodine}

\subsection{Measuring molecular alignment by ion imaging}
\label{sec:ion-imaging}

The basic experimental observable that allows us to infer information about the spatial orientation of the molecules is 2D velocity images of the recoil ions from Coulomb explosion induced by the probe pulse. For the studies on isolated \ce{I_2} molecules \ce{I^+} ions are detected whereas for \ce{I_2} in He droplets we choose to detect \ce{IHe^+} ions~\cite{braun_photodissociation_2007-1} because they can only be produced from molecules inside the droplets and not from those isolated molecules that manage to effuse from the pick-up region to the interaction zone in the target chamber. As such the \ce{IHe+} signal is ‘background-free’.  For the measurement on DIB and DBB in He droplets, presented in \autoref{sec:res-dib-dbb}, we recorded images of the bare atomic ions, \ce{I^+} and \ce{Br^+}. Here the background contribution from ionization of isolated molecules effusing into the target region from the pick-up cell had to be corrected for in order to obtain accurate values for the degree of alignment.

\begin{figure}%
\includegraphics[width=\columnwidth]{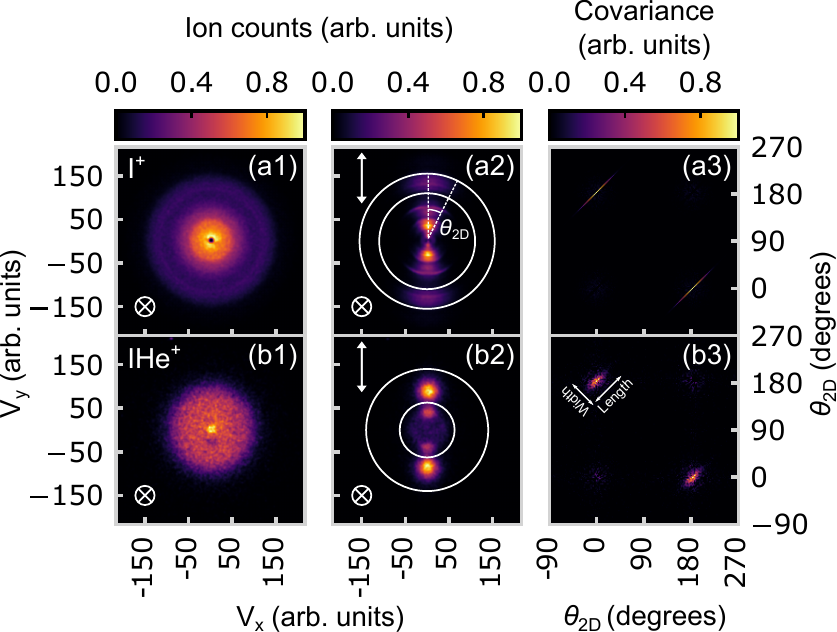}
\caption{(a1),(a2): \ce{I^+} ion images from Coulomb exploding isolated \ce{I_2} molecules with the probe pulse only and with the alignment pulse included. The images represent the detection of the ion velocities, v$_\text{x}$, v$_\text{y}$ in the detector plane. (a3): Angular covariance map of the image in (a2).
(b1),(b2): \ce{IHe^+} ion images from Coulomb exploding \ce{I_2} molecules in He droplets with the probe pulse only and with the alignment pulse included. (b3): Angular covariance map of the image in (b2). The polarization directions of the alignment pulse (vertical: $\updownarrow$) and probe pulse (perpendicular to the detector plane: $\otimes$) are shown on the ion images. White circles indicate the radial ranges used for calculating $\costhetwod$ and the angular covariance maps. For the images in the middle column the probe pulse was sent at $t$ = 0 and I$_\text{align}$ = 0.83 TW/cm$^2$. The direction of the length and width is illustrated in (b3). }
\label{fig:i2-ion-images}%
\end{figure}

The probe pulse is polarized perpendicular to the detector plane. With this probe pulse polarization the \ce{I^+} and \ce{IHe+} images are circularly symmetric when the molecules only interact with the probe pulse [\autoref{fig:i2-ion-images}(a1) and (b1)].  When the alignment pulse is included and the probe pulse is synchronized to its peak ($t$ = 0) the \ce{I^+} and \ce{IHe+} ions localize along its polarization direction at certain times. Examples of this are displayed in \autoref{fig:i2-ion-images}(a2) and (b2). These images show that the molecules are 1D aligned. To characterize the alignment we determine cos$^2\theta_{2D}$ for each ion hit on the detector in the radial ranges, indicated by the white circles in \autoref{fig:i2-ion-images}(a2) and \autoref{fig:i2-ion-images}(b2). Here $\theta_{2D}$ is the angle between the polarization axis of the alignment pulse and the projection of the recoil vector of an \ce{I^+} or \ce{IHe+} ion on the detector plane [see \autoref{fig:i2-ion-images}(a2)]. The ensemble average, $\costhetwod$, of all \ce{I^+} or \ce{IHe+} ions in the respective radial ranges, constitutes the quantitative measure for the degree of alignment.

In the case of the isolated \ce{I_2} molecules the radial range was chosen to select those \ce{I^+} ions that stem from double ionization of \ce{I_2} molecules and subsequent fragmentation into two \ce{I^+} ions:  \ce{I_2^{2+}} $\rightarrow$ \ce{I^+} + \ce{I^+}. For this Coulomb explosion channel momentum conservation implies that the two ions should recoil with a relative angle of 180$\degree$. Therefore, we expect that the axial recoil approximation is met to a very high degree, i.e. that the emission direction of an \ce{I^+} ion should essentially be identical to the orientation of the internuclear axis of the parent molecule at the instant the probe pulse interacts with the molecules. Experimentally, this is identified from the angular covariance map~\cite{hansen_control_2012} depicted in \autoref{fig:i2-ion-images}(a3) determined from the \ce{I^+} ions in the selected radial range of the image in \autoref{fig:i2-ion-images}(a2). Two narrow diagonal lines centered at (\SI{0}{\degree}, \SI{180}{\degree}) and (\SI{180}{\degree}, \SI{0}{\degree}) stand out in the covariance map showing that the emission direction of an \ce{I^+} ion is strongly correlated with another \ce{I^+} ion departing in the opposite direction. This identifies the ions as originating from the \ce{I^+}-\ce{I^+} channel. As discussed in detail in Ref.~\cite{christensen_deconvoluting_2016} the length of the oblong covariance islands is a measure of the distribution of the molecular axes, i.e. the degree of alignment, whereas the width is a measure of the degree of axial recoil. The observation of a width of $\sim$ 1$\degree$ shows that the axial recoil approximation is a good approximation.

In the case of the molecules inside He droplets the angular covariance map also shows two oblong islands centered at  (\SI{0}{\degree}, \SI{180}{\degree}) and (\SI{180}{\degree}, \SI{0}{\degree}) -– see \autoref{fig:i2-ion-images}(b3). This identifies the ions as originating from double ionization of the \ce{I_2} molecules and fragmentation into a pair of correlated \ce{IHe+} ions. While the length of the covariance islands is still small, thus showing that the molecules are well aligned, the width, $\sim$ 11$\degree$, is much larger than the case for the isolated molecules. We believe the increased width is due to distortion of the ion trajectories from collisions with He atoms on the way out of the droplet. The angular distribution of the \ce{IHe+} ions is, therefore, wider than the distribution of the molecular axes and represents an underestimate of the real degree of alignment.  As discussed in \autoref{sec:i2-align-intensity} we are able to deconvolute the effect of the deviation from axial recoil in the probe process using analysis of the angular covariance maps~\cite{christensen_deconvoluting_2016}.

\subsection{Dependence of alignment on intensity}
\label{sec:i2-align-intensity}

\begin{figure}%
\includegraphics[width=\columnwidth]{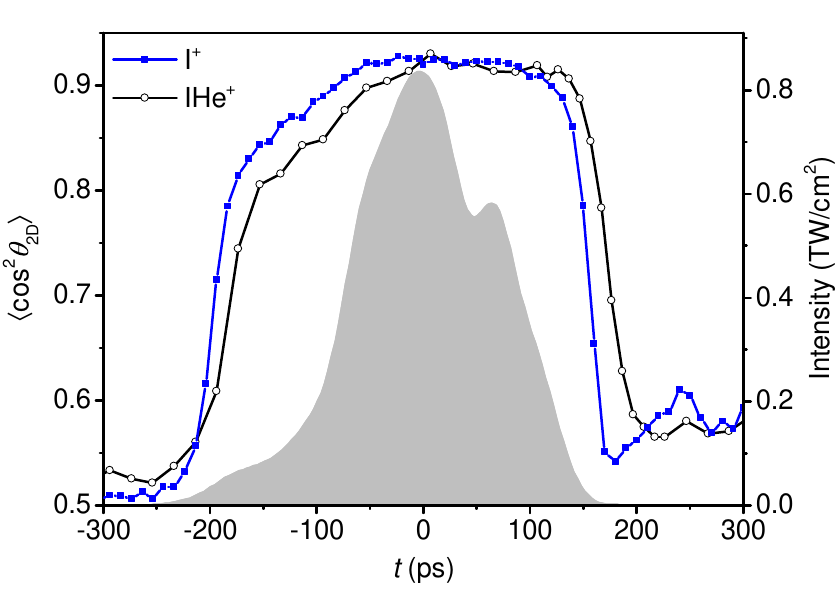}
\caption{The degree of alignment, quantified by $\costhetwod$, as a function of time for iodine molecules both solvated inside helium droplets (open black circles) and isolated in the gas phase (filled blue squares) recorded over the duration of the alignment pulse (shaded grey area). The intensity of the alignment pulse is shown on the right vertical axis.}
\label{fig:alignmentpulse}%
\end{figure}

In the current work we aim to measure the strongest alignment as a function of intensity. If the alignment field is turned on slower than the rotational period of the molecule under study, the alignment process proceeds in the adiabatic regime and the highest degree of alignment is expected to occur when the alignment field reaches its maximum -- according to past theory and experiments on isolated molecules~\cite{ortigoso_time_1999,sakai_controlling_1999,torres_dynamics_2005,trippel_strongly_2014}. In the current study the alignment pulse is turned on in about \SI{150}{ps} -- see the intensity profile of the pulse in \autoref{fig:alignmentpulse}. This is faster than the rotational period of isolated \ce{I_2} molecules ( = $h/2B$ = 446 ps, where B is the rotational constant in SI energy units) implying that the alignment dynamics may differ from strict adiabatic behaviour~\cite{trippel_strongly_2014}. To identify the instant where the alignment is strongest we, therefore, measured $\costhetwod$ as a function of time. The result for the isolated \ce{I_2} molecules, represented by the blue curve (filled squares) in \autoref{fig:alignmentpulse}, demonstrates that $\costhetwod$ reaches its peak value at $t$ = 0 ps, i.e. at the time of the highest intensity just as in the adiabatic regime of alignment. Likewise, for \ce{I_2} molecules in He droplets the strongest alignment also occurs at the peak of the laser pulse as can be seen from the black curve (open circles) in \autoref{fig:alignmentpulse}.  The data points show, however, significant deviations from adiabatic behavior during and after the trailing part of the laser pulse. Notably the molecules do not return to free rotor states characterized by random orientation with $\costhetwod$ = 0.50 after the pulse has turned off. Rather $\costhetwod$ stays above 0.50 and measurements at times out to 1500 ps reveal revival structures for both the isolated molecules and the molecules inside the droplets. These nonadiabatic effects are not of importance for the current studies but will be reported in a different paper~\cite{Shepperson-2017-nonadiabatic}.

\begin{figure}%
\includegraphics[width=\columnwidth]{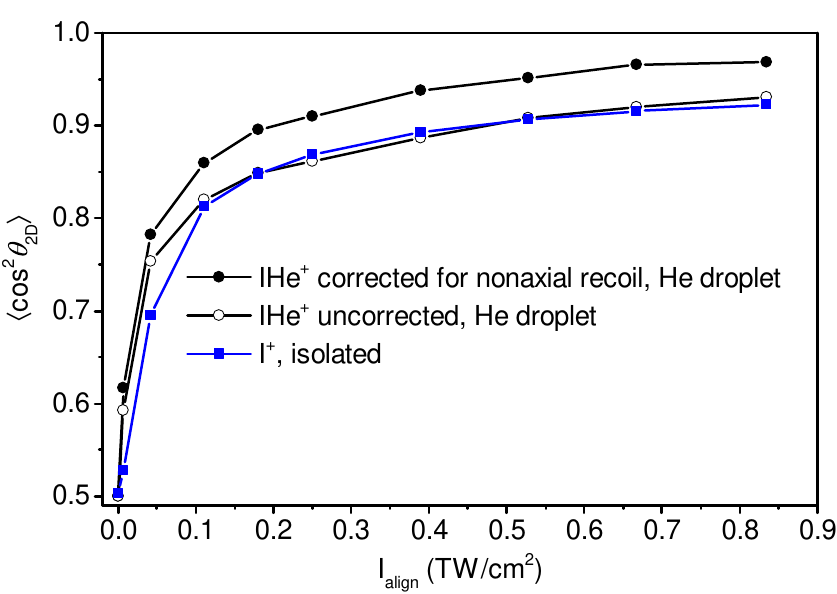}
\caption{The degree of alignment measured at the peak of the alignment pulse as a function of laser intensity for \ce{I^+} ions from isolated molecules (filled blue squares) and \ce{IHe^+} ions (open black circles) from molecules inside helium droplets. The degree of alignment for \ce{IHe^+} ions after correcting for nonaxial recoil is also shown (filled black circles).}
\label{fig:gascomp}%
\end{figure}

\Autoref{fig:gascomp} shows $\costhetwod$, obtained at $t$ = 0 ps, as a function of the intensity of the alignment pulse, I$_\text{align}$. The values of $\costhetwod$ determined directly from the ion images are represented by the blue (filled squares) curve (isolated \ce{I_2}) and the black (open circles) curve (\ce{I_2} in He droplets). Both curves rise gradually from 0.50, the value corresponding to randomly oriented molecules, at I$_\text{align}$ = 0 W/cm$^2$ to $\sim$ 0.92 at the highest intensity used. We expect the $\costhetwod$ values of the isolated molecules to represent an accurate measurement of the true degree of alignment due to the essentially perfect axial recoil of the \ce{I^+} fragments. For the molecules in He droplets the deviation of axial recoil, explained in \autoref{sec:ion-imaging}, implies that the true degree of alignment is underestimated. Using a recently developed method based on analysis of the correlation between the \ce{I^+} fragment ions (employing the angular covariance maps) it is possible to correct the $\costhetwod$ measured for the effect of nonaxial recoil~\cite{christensen_deconvoluting_2016}. The result of this correction method is represented by the black curve (filled circles) in \autoref{fig:gascomp}. Comparing this curve to the blue curve (filled squares) it is clear that $\costhetwod$ is significantly higher for \ce{I_2} in He droplets than for isolated \ce{I_2} molecules at all intensities.

\subsection{Comparison of experimental results to calculations}
\label{sec:i2-comparison}

\begin{figure}%
\includegraphics[width=\columnwidth]{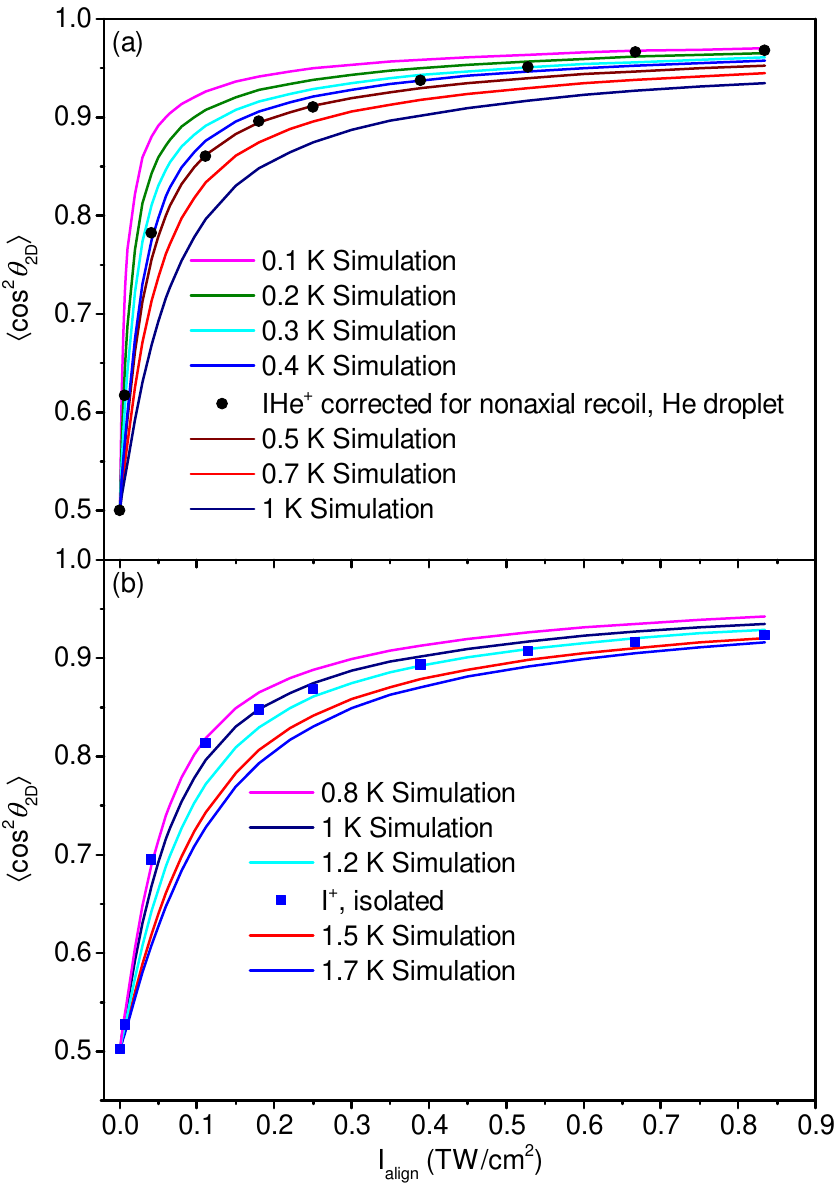}
\caption{$\costhetwod$ for (a) \ce{I2} molecules in He droplets and (b) isolated \ce{I2} molecules calculated as a function of the alignment pulse intensity for different rotational temperatures. The black filled circles in (a), representing the experimental results (He droplets), are the same as the black filled circles in \autoref{fig:gascomp} and the filled blue squares (experimental results for isolated molecules) are the same as the filled blue squares in \autoref{fig:gascomp}.}
\label{fig:calc-I2}%
\end{figure}

According to the angulon theory described in \autoref{sec:theory} the conditions of the experiment falls in a regime where the alignment of the \ce{I_2} molecules inside He droplets can be approximated as gas-phase alignment, provided the molecules are characterized by an effective rotational constant, $B^\ast$. Consequently, we calculated $\costhetwod$ as a function of intensity at different temperatures. The calculations were done using a recently developed program~\cite{sondergaard_nonadiabatic_2017} that solves the time-dependent rotational Schr\"{o}dinger equation for isolated molecules and outputs both $\langle\cos^2\theta\rangle$, the commonly used metric in theoretical descriptions of alignment ($\theta$: the polar Euler angle between the molecular axis and the alignment pulse polarization axis) and $\costhetwod$, which allows a direct comparison to the experimental results. The calculations used the gas-phase polarizability anisotropy, the experimentally determined focal spot-sizes of the alignment and the probe beam, and the value of $B^\ast$ theoretically determined previously~\cite{Shepperson-2017-nonadiabatic-short}. The calculated results along with the experimental results are displayed in \autoref{fig:calc-I2}(a).

It appears that the experimental data points at all intensities fall within the range confined by the calculated 0.2 K and 0.5 K curves, and on average lie close to the calculated 0.4 K curve. In fact, we believe the agreement between the data points and the 0.4 K curve would be even better if the former could be corrected for the alignment-dependent ionization efficiency of the probe pulse: The probe pulse ionizes, and thus Coulomb explodes, the molecules aligned parallel (perpendicular) to its polarization axis most (least) efficiently. At intermediate degrees of alignment such as for the three data points between 0.1 and 0.3 TW/cm$^2$ $\costhetwod$ will be underestimated because the best aligned molecules are least efficiently probed. A correction of this probe pulse selectivity would increase $\costhetwod$ for the three points and bring them closer to the 0.4 K curve. The agreement between the theoretical curve for 0.4 K and the experimental results strongly indicate that alignment of \ce{I_2} molecules in He droplet, induced by the \SI{160}{ps} pulse, behaves like a sample of gas-phase molecules characterized by $B^\ast$ and with the temperature of the He droplet, i.e. 0.4 K. This low temperature is the reason that a $\costhetwod$ value as high as 0.96 is obtained at an intensity of 0.83 TW/cm$^2$.

A similar comparison of the experimental data point for the isolated molecules and calculated results is shown in \autoref{fig:calc-I2}(b). It can be seen that the data points (besides the point at the highest intensity) fall within the range confined by the calculated 0.8 K and 1.2 K curves. A temperature around 1 K is consistent with previous estimates of the rotational temperature for molecular beams produced by an Even Lavie valve operated under similar conditions~\cite{filsinger_quantum-state_2009}.

The analysis in this section shows that the \ce{I_2} molecules embedded in He droplets align better than the isolated molecules because the 0.4 K rotational temperature inside the droplets is lower than that of a molecular beam - even when produced by a high-pressure supersonic expansion.

\section{Results for \DIBlong and \DBBlong molecules}
\label{sec:res-dib-dbb}

\begin{figure}%
\includegraphics[width=\columnwidth]{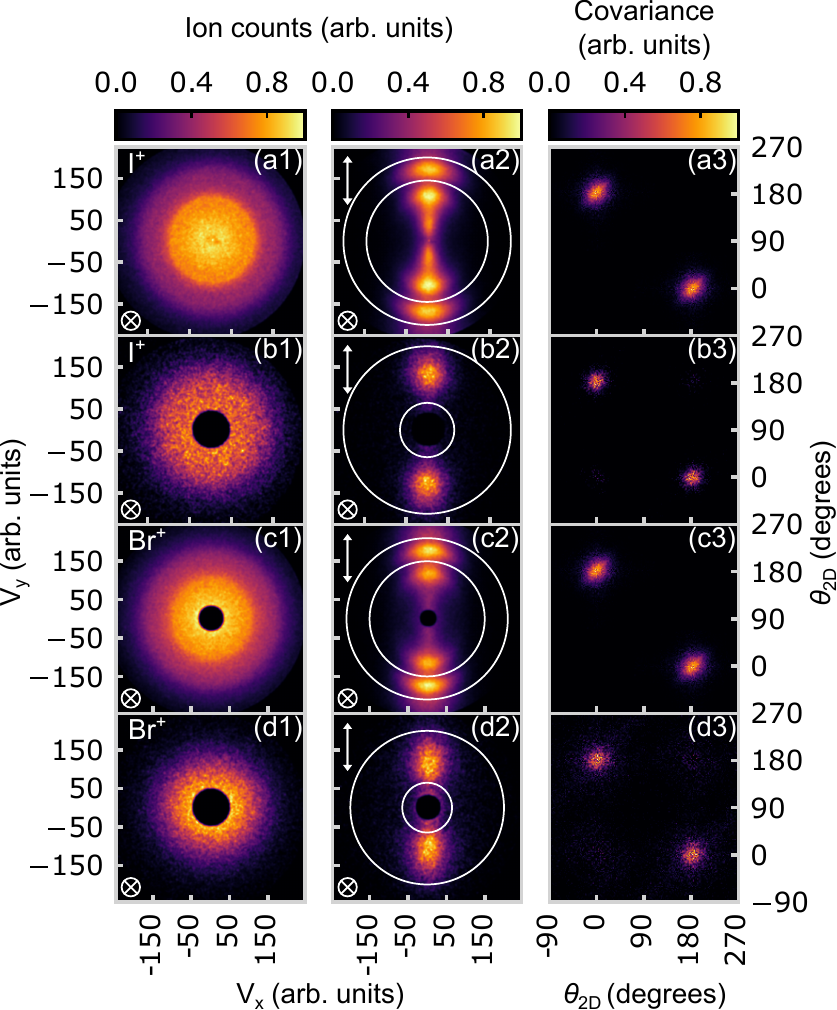}
\caption{ (a1),(a2): \ce{I^+} ion images from Coulomb exploding isolated DIB molecules with the probe pulse only and with the alignment pulse included. (a3): Angular covariance map of the image in (a2). (b1)-(b3): As (a1)-(a3) but for DIB molecules in He droplets. (c1),(c2): \ce{Br^+} ion images from Coulomb exploding isolated DBB molecules with the probe pulse only and with the alignment pulse included. (c3): Angular covariance map of the image in (c2). (d1)-(d3): As (c1)-(c3) but for DBB molecules in He droplets. The polarization directions of the alignment pulse (vertical: $\updownarrow$) and probe pulse (perpendicular to the detector plane: $\otimes$) are shown on the ion images. The white circles indicate the radial ranges used for calculating $\costhetwod$ and the angular covariance maps. For the images in column 2 the probe pulse was sent at t = 0 and I$_\text{align}$ = 0.83 TW/cm$^2$ }
\label{fig:dibdbb-images}%
\end{figure}

We also conducted measurements on DIB and DBB molecules to explore the alignment of more complex molecules. The procedure for the experiments was the same as  that used for the iodine molecules. Examples of ion images are given in \autoref{fig:dibdbb-images}. The results for the isolated molecules are depicted in row (a) and (c). Panels (a2) and (c2) are ion images recorded at the highest intensity used at the peak of the alignment pulse. The strong localization of the ions along the polarization axis shows that the molecules are tightly 1D aligned, i.e. that the I-I (Br-Br) axis is tightly confined along the alignment laser polarization axis. Again the pronounced islands in the covariance maps calculated for the ions between the white circles show that the ions stem from Coulomb explosion producing a pair of \ce{I^+} ions (\ce{Br^+} ions). In fact, previous analysis for DIB molecules shows that the primary contribution to the \ce{I^+} ions is the fragmentation pathway:
\ce{C6H4I2^3+} $\rightarrow \ce{I+} + \ce{C6H4+} + \ce{I+}$~\cite{christiansen_laser-induced_2016}. Unlike the case for the \ce{I_2} molecules [\autoref{fig:i2-ion-images}(a3)] the covariance islands have a finite width [panel (a3) and (c3)], i.e. there are deviations from perfect axial recoil even for the isolated molecules. The results for the DIB and DBB molecules in He droplets are displayed in row (b) and (d). Panels (b2) and (d2) show that the molecules strongly 1D align and panels (b3) and (d3) show that the Coulomb explosion process, creating the \ce{I^+} and \ce{Br^+} ions, is subject to deviation from axial recoil.

\begin{figure}%
\includegraphics[width=\columnwidth]{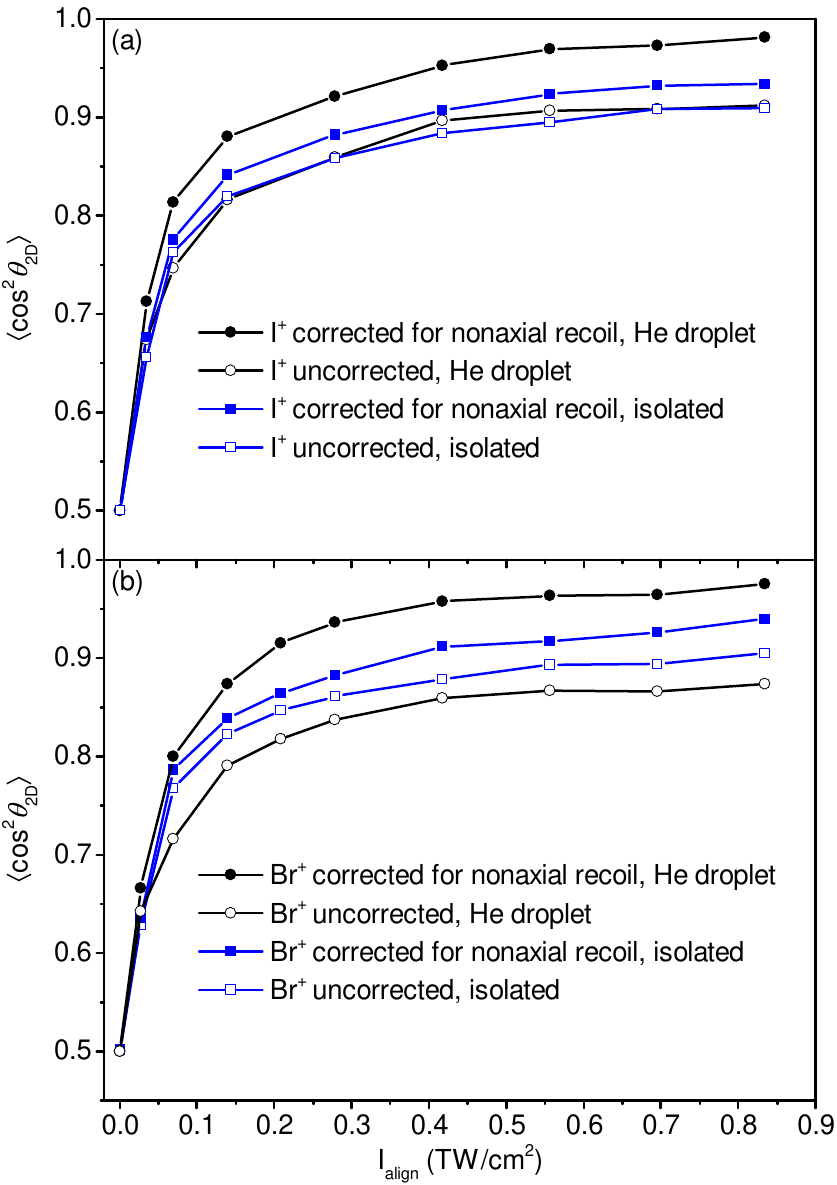}
\caption{The degree of alignment measured at the peak of the alignment pulse as a function of laser intensity for DIB molecules [panel (a)] and DBB molecules [panel (b)] both in gas phase and inside He droplets. The figures show both the $\costhetwod$ values obtained directly from the ion images ('uncorrected') and after deconvolution of the nonaxial recoil ('corrected') -- identified by the legends.}
\label{fig:dibdbb-degree-alignment}%
\end{figure}

\Autoref{fig:dibdbb-degree-alignment}(a) shows $\costhetwod$, obtained at $t$ = 0 ps, as a function of the intensity of the alignment pulse for DIB molecules. The blue (open squares) curve (isolated DIB) and the black (open circles) curve (DIB in droplets) are the $\costhetwod$ values obtained directly from the \ce{I^+} images. It is seen that the two curves are very similar starting at 0.50 and ending at 0.91 at the highest intensity. Correction for the non-axial recoil in the Coulomb explosion gives the blue (filled squares) and the black (filled circles) curve, respectively. In both cases the curves are lifted towards higher $\costhetwod$ values but the effect is most significant for the DIB molecules in the droplets resulting in a maximum $\costhetwod$ value in excess of 0.96. The reason for the larger correction for the molecules in He droplets is that the deviation from axial recoil is larger here than for isolated molecules. This can be seen by comparing the transverse width of the covariance islands in \autoref{fig:dibdbb-images}(a3) and (b3). For the isolated molecules the width is $20\degree$ and for the molecules in He droplets the width is $29\degree$ using the convention of Ref.~\cite{christensen_deconvoluting_2016}.

Very similar observations are made for the DBB molecules. The $\costhetwod$ curves determined directly from the \ce{Br^+} ion images show a slightly higher degree of alignment for the isolated molecules (blue, open squares, curve) compared to the molecules in He droplets (black, open circles, curve). Since the deviation from perfect axial recoil is larger in the droplet case (transverse width of covariance islands is $33\degree$ for the molecules in droplets and $20\degree$ for the isolated molecules) the correction process leads to a higher lying $\costhetwod$ curve for the molecules in He droplets (black, filled circles, curve) compared to the isolated molecules (blue, filled squares, curve). As for the case of DIB the highest $\costhetwod$ value for the DBB molecules in the droplets exceeds 0.96. The correction procedure for the nonaxial recoil is slightly more involved for the DIB and DBB molecules in He droplets because there is a background of \ce{I^+} and \ce{Br^+} from the isolated molecules that effuse from the doping region into the target region. In the appendix we briefly account for how we correct for the effusive background.

The conclusion from \autoref{fig:dibdbb-degree-alignment} is that for both DIB and DBB molecules the degree of alignment at the peak of the alignment pulse is significantly higher for the molecules in the He droplets compared to the isolated molecules. This is similar to the findings for the \ce{I_2} molecules and like that case we believe the principal reason is the lower rotational temperature of the molecules inside the He droplets compared to that of the molecular beam. Our program for calculating the degree of alignment only applies to linear and symmetric top molecules and not to the asymmetric tops DIB and DBB. Consequently, we were not able to test if the experimental $\costhetwod$ curve followed a $\costhetwod$ curve calculated for 0.4 K.


\section{Conclusion}
\label{sec:conclusion}

We used angulon theory, describing rotating molecules in superfluid helium droplets, to argue that the mechanism of laser-induced alignment of molecules in He droplets is similar to that of isolated molecules, described by an effective rotational constant $B^\ast$, provided the duration (or more precisely the turn-on and turn-off time) of the laser pulse is shorter than the time scale of He excitations ($\sim$ 7 ps) and of He-molecule interactions ($\sim$ 14 ps for I$_2$-He).


Experimentally we measured alignment, induced by a 160 ps long pulse, of \ce{I_2} molecules inside He droplets. The highest degree of alignment, attained at the peak of the pulse, was determined as a function of intensity and compared to calculations based on numerical solution of the rotational Schr\"{o}dinger equation for gas phase molecules, characterized by $B^\ast$. The very good agreement between the measurements and the calculated results for a temperature of 0.4 K -- equal to that of the droplet -- confirms the theoretical prediction of a gas-phase-like alignment mechanism for molecules embedded in He droplets.

In the near-adiabatic alignment regime studied here the rotational temperature is the main parameter determining the degree of alignment for a given molecule and a given intensity~\cite{seideman_dynamics_2001}. As a consequence very high degrees of alignment can be achieved due to the 0.4 K temperature. The current work demonstrated that for iodine, \DIBlong and \DBBlong molecules. Any molecule embedded into He droplets is expected to obtain a temperature of 0.4 K and therefore, we expect that similar high degrees of alignment can be achieved for an extensive range of molecules. For molecules with moments of inertia much larger than \DIBlong it may necessary to apply longer alignment pulses to ensure (near-) adiabatic conditions. This can be realized by standard pulse-stretching in a pair of diffraction gratings~\cite{trippel_strongly_2013}. Furthermore, it should be possible to convert the strong alignment to field-free alignment, lasting several picoseconds, by truncating the alignment pulse at its peak with the plasma shutter technique~\cite{underwood_switched_2003,takei_laser-field-free_2016} or with spectral filtering~\cite{Chatterley-3D}.

\acknowledgements

We acknowledge support from the European Research Council-AdG (Project No. 320459, DropletControl), the Villum Foundation, and the Austrian Science Fund (FWF), grant Nr. P29902-N27.


\appendix*\section{Background correction for effusive molecules}
\label{sec:Appendix}

\begin{figure}%
\includegraphics[width=\columnwidth]{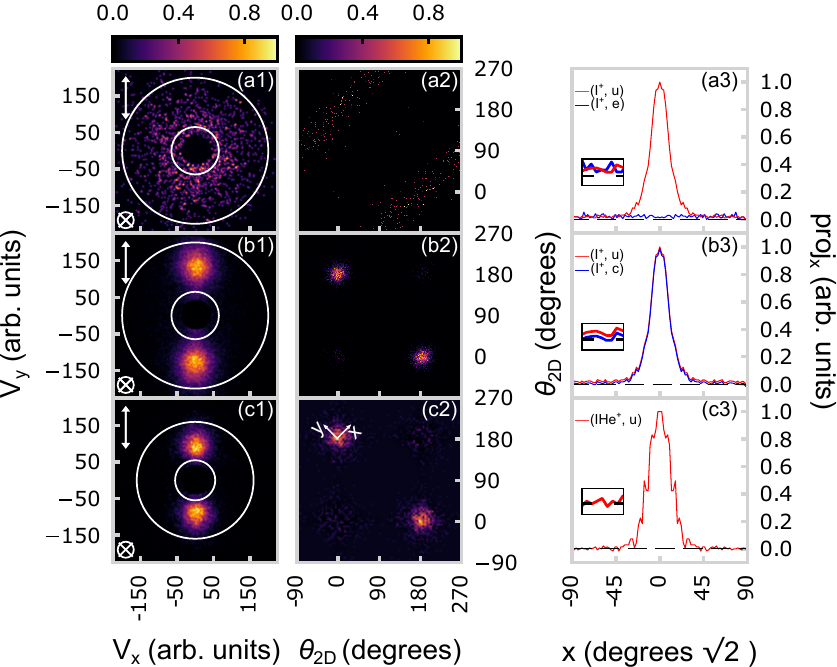}
\caption{
(a1): \ce{I^+} ion image from effusive DIB molecules, (b1): \ce{I^+} ion image from DIB molecules in helium droplets and (c1): \ce{IHe^+} ion image from DIB molecules in helium droplets -- all obtained with the probe and the alignment pulses included. (a2),(b2),(c2): Angular covariance maps of the ion images shown in (a1),(b1),(c1). Panel (c2) shows the axes used to determine the width (y) and the length (x) of the covariance islands. (a3): Projection of the covariance island shown in (a2) (\ce{I^+}, e) and (b2) (\ce{I^+}, u) onto the x axis. (b3): Projection of the covariance island shown in (b2) onto the x axis with (\ce{I^+}, c) and without (\ce{I^+}, u) correction for a constant y offset. (c3): Projection of (c2) onto the x axis without correction for a y offset (\ce{IHe^+}, u). In panels (a3)-(c3) the black dashed horizontal line shows the zero level on the y-axis and the inserts show an expansion of this region. }
\label{fig:backgroundsub}%
\end{figure}

For the studies on DIB and DBB molecules inside helium droplets \ce{I^+} and \ce{Br^+} ions, respectively, were detected. Therefore, any \ce{I^+} or \ce{Br^+} ions resulting from ionization of effusive molecules not being captured by the liquid nitrogen traps, will be present in the ion images and in the angular covariance maps obtained for the He droplet measurements. Without correcting for the contribution from effusive molecules the degree of alignment will be underestimated.

In this work a background subtraction procedure was implemented to correct for the contribution of the effusive molecules. \Autoref{fig:backgroundsub}(a1) shows the \ce{I^+} ion image from effusive \DIB molecules, i.e. with the helium droplet beam blocked, and the doping optimized for the pick-up of a single \DIB molecule per droplet with both the alignment and the probe pulse included. There is essentially no localization of the ions along the  polarization axis of the alignment pulse. Instead the ions are characterized by an almost random orientation with $\costhetwod$ = 0.52. In the corresponding angular covariance map two faint diagonal lines with a uniform distribution are observed. This confirms that the effusive molecules are randomly aligned as there is no preference for the emission direction of \ce{I^+} ions. The random orientation is a result of the fact that the effusive molecules have a rotational temperature of 295 K determined by the temperature in the laboratory.

\Autoref{fig:backgroundsub}(a3) shows the projection of the covariance islands in panel (a2) and (b2) onto the x axis, the blue (\ce{I^+}, e) and red (\ce{I^+}, u) curve respectively. The intensity was scaled such that the maximum intensity of (b2) was 1 and the same scaling factor was applied to (a2). What is clear from panel (a3) is that the signal from the effusive molecules (the blue curve) has the same y offset as the signal from the aligned DIB molecules inside helium droplets (the red curve). This identifies the y offset of the red curve as resulting from the effusive molecules and, thus, this can be removed simply by subtracting the blue curve from the red curve.

Due to the low signal and long data acquisition times for molecules solvated in helium droplets it is undesirable to have to record an effusive only image for every data point to use for the background subtraction. However, a more convenient method is to fit the projection of the angular covariance map for the droplet solvated molecules onto the x axis to a Gaussian and subtract the y offset obtained from the fit. This method was used to correct the data shown in (b3) -- the blue trace (\ce{I^+}, c). To confirm that this offset was the result of effusive molecules the results from gating on the \ce{IHe^+} ion is shown in row (c). From the projection of (c2) onto the x axis there is no y offset, which is expected given that this ion species cannot be formed without the presence of the helium droplet and no effusive molecules will be present when gating on \ce{IHe^+}.

%

\end{document}